\documentclass[preprintnumbers,showpacs,aps,prd,epsf,nofootinbib,twocolumn]{revtex4}
\usepackage{latexsym,amsmath}
\usepackage{amssymb}
\usepackage{color}
\usepackage{graphicx,bm}% Include figure files
\usepackage{dcolumn}    % Align table columns on decimal point
\usepackage{subfigure}  % sub labeling of fingures
\definecolor{red  }{rgb}{1,0,0}
\definecolor{blue }{rgb}{0,0,1}
\definecolor{green}{rgb}{0,1,0}

\begin{document}
\thispagestyle{empty}

\title{ 
Dynamics of colliding branes and black brane production
}

\author{Yu-ichi Takamizu$^{1}$}
\email{takamizu_at_gravity.phys.waseda.ac.jp}
\author{Hideaki Kudoh$^{2,3}$}
%% \email{kudoh_at_utap.phys.s.u-tokyo.ac.jp}
\author{Kei-ichi Maeda $^{1}$}
%% \email{maeda_at_gravity.phys.waseda.ac.jp}

\address{
$^{1}$Department of Physics, Waseda University,
Okubo 3-4-1, Shinjuku, Tokyo 169-8555, Japan
}
\address{
$^{2}$Department of Physics, UCSB, Santa Barbara, CA 93106, USA
}
\address{
$^{3}$Department of Physics, The University of Tokyo, 
Tokyo 113-0033, Japan,
}

\preprint{UTAP-574}

\begin{abstract}
We study the dynamics of colliding domain walls including self-gravity. 
The initial data is set up by applying a BPS domain wall in five-dimensional supergravity, and we evolve the system determining the final outcome of collisions. 
After a collision, a spacelike curvature singularity covered by a horizon is formed in the bulk, resulting in a black brane with trapped domain walls. 
This is a generic consequence of collisions, except for non-relativistic weak field cases, in which the walls pass through one another or multiple bounces take place without singularity formation. 
These results show that incorporating the self-gravity drastically changes 
a naive picture of colliding branes. 
\end{abstract}

\maketitle

%----------------------------------
\section{Introduction}
%----------------------------------

It is known that primordial black holes and domain walls may have been produced in the early universe through the physical process of the collapse of cosmological density perturbations and the series of phase transitions during the cooling phase of universe. 
On the other hand, black holes and domain walls ({also} known as branes) also play an important role in string theory as fundamental constituents. 
In addition, according to M-theory, branes are of particular relevance to cosmology: branes are free to move in a bulk space, and they may approach and collide, causing the big bang/crunch or an inflation on branes~\cite{Khoury:2001wf}.

In view of the phenomenological relevance, understanding how the domain walls/branes interact dynamically is an important problem, and {more knowledge in this area} could help in clarifying many issues regarding the early universe.  
In the past few years much attention have been paid to understanding the dynamics of domain walls and bubbles~(e.g.,~\cite{Takamizu:2006gm,Shinkai:1993vk} and \cite{Blanco-Pillado:2003hq}). 
In particular, the interaction between black holes and domain walls has been the subject of study.    
Nevertheless, even more fundamental processes like collision, recoil, and reconnection of branes are less understood.

The collision and recoil of domain walls in the cosmological context described above was studied in \cite{Takamizu:2006gm}, where a reheating mechanism via particle productions was discussed within a toy model. 
In this paper, we consider the problem from a different perspective. 
The collision of domain walls/branes is a violent phenomenon, and, as partially observed in our previous study, a spacetime singularity might appear through a collision. 
If this is the case, a low-energy description of colliding branes breaks down at some point,  
impling a complete loss of predictability, without the complete theory of quantum gravity.

We investigate the process of collision using a BPS domain wall in five-dimensional supergravity, and our main goal is to determine the final outcome of the kink-anti-kink collisions including self-gravity.
As we will see, singularity formation is a generic consequence of collisions. 
However, the singularity is spacelike and hidden inside the horizon. 
The horizon  {extends in a} spatially flat direction along the brane so that a black brane is produced through the collision. 
To clarify and to provide further examples of black brane production, we will also study collisions using another model of domain walls.

\vspace{-0.5cm}
%----------------------------------
\section{Brane dynamics}
%----------------------------------

%----------------------------------
\subsection{Model of wall collisions}
%----------------------------------

%%%%%%%%%%%%%%%%%%%%%%%%%%%%%%%%%%%%%%%%%%%%%%
\begin{figure*}[bt]
\subfigure[$\delta_K=\delta_A=1, ~{L}_K={L}_A=0.18$ ]{
\includegraphics[width=4.2cm,angle=-90]{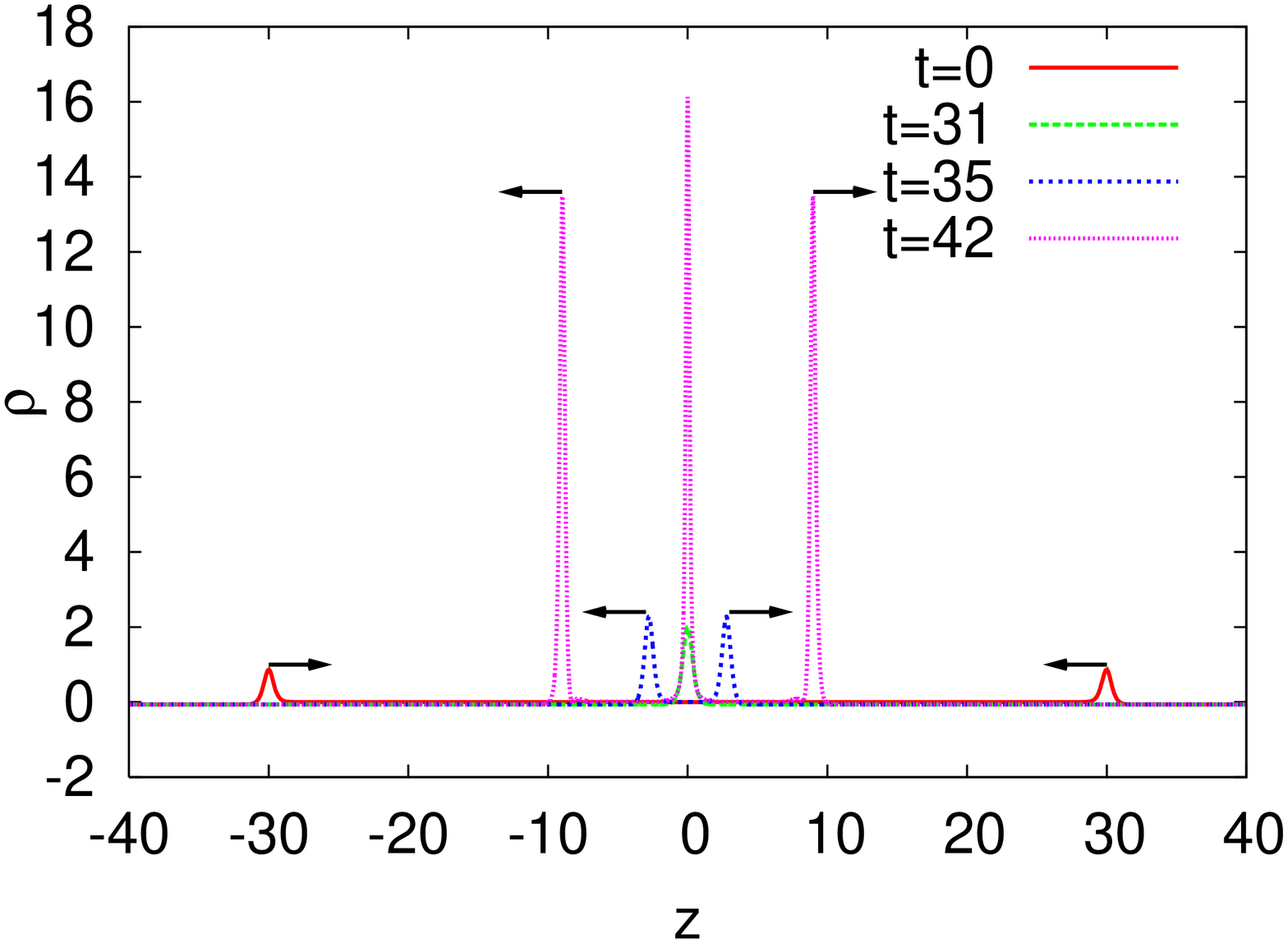}
\label{fig:rho vs z:A}
}
\hspace{-0.7cm}
\subfigure[$\delta_K=1$ and $\delta_A=2$]{
\includegraphics[width=4.2cm,angle=-90]{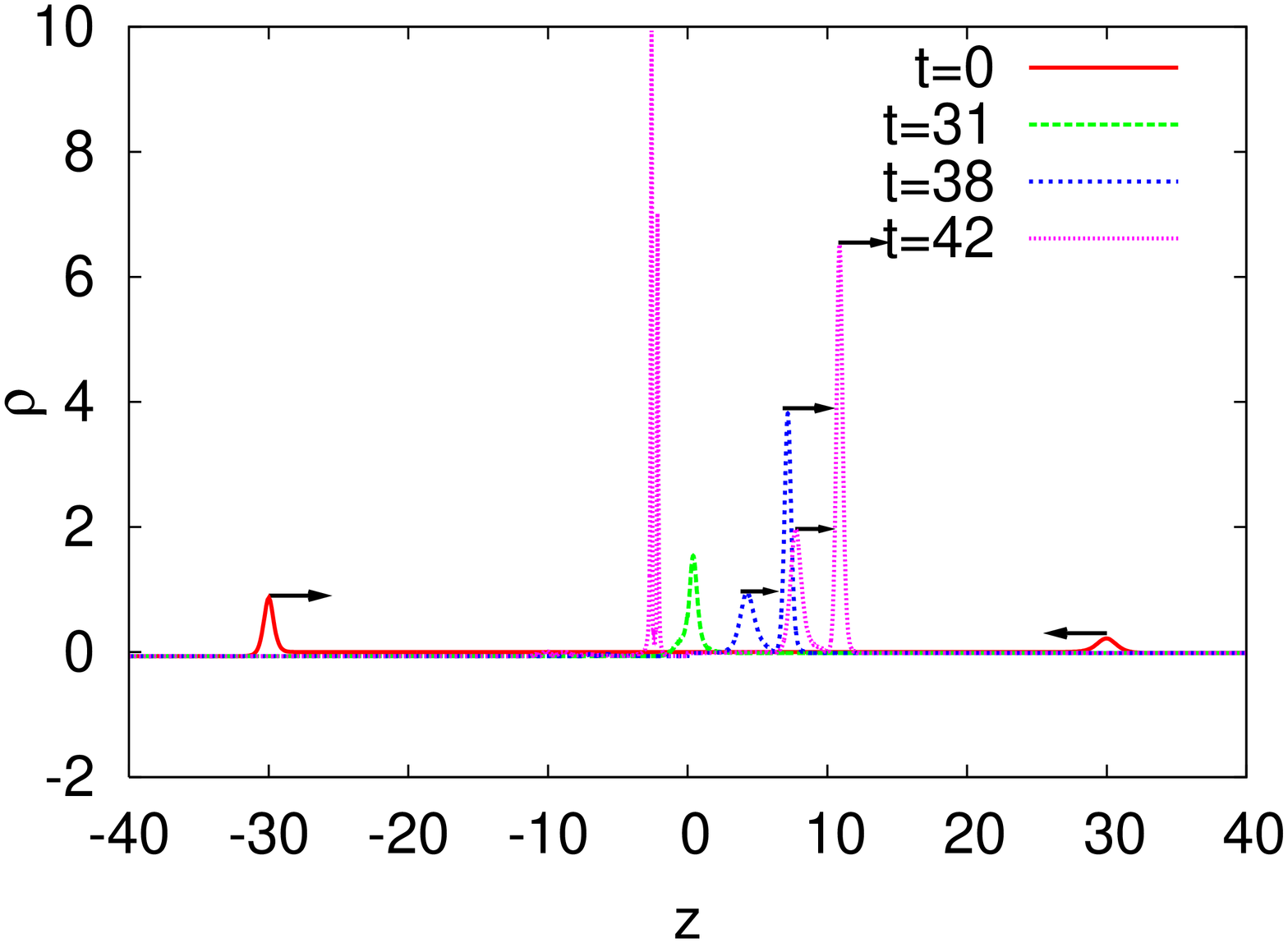}
\label{fig:rho vs z:B}
}
\hspace{-0.7cm}
\subfigure[${L}_K=0.18$ and ${L}_A=0.22$]{
\includegraphics[width=4.2cm,angle=-90]{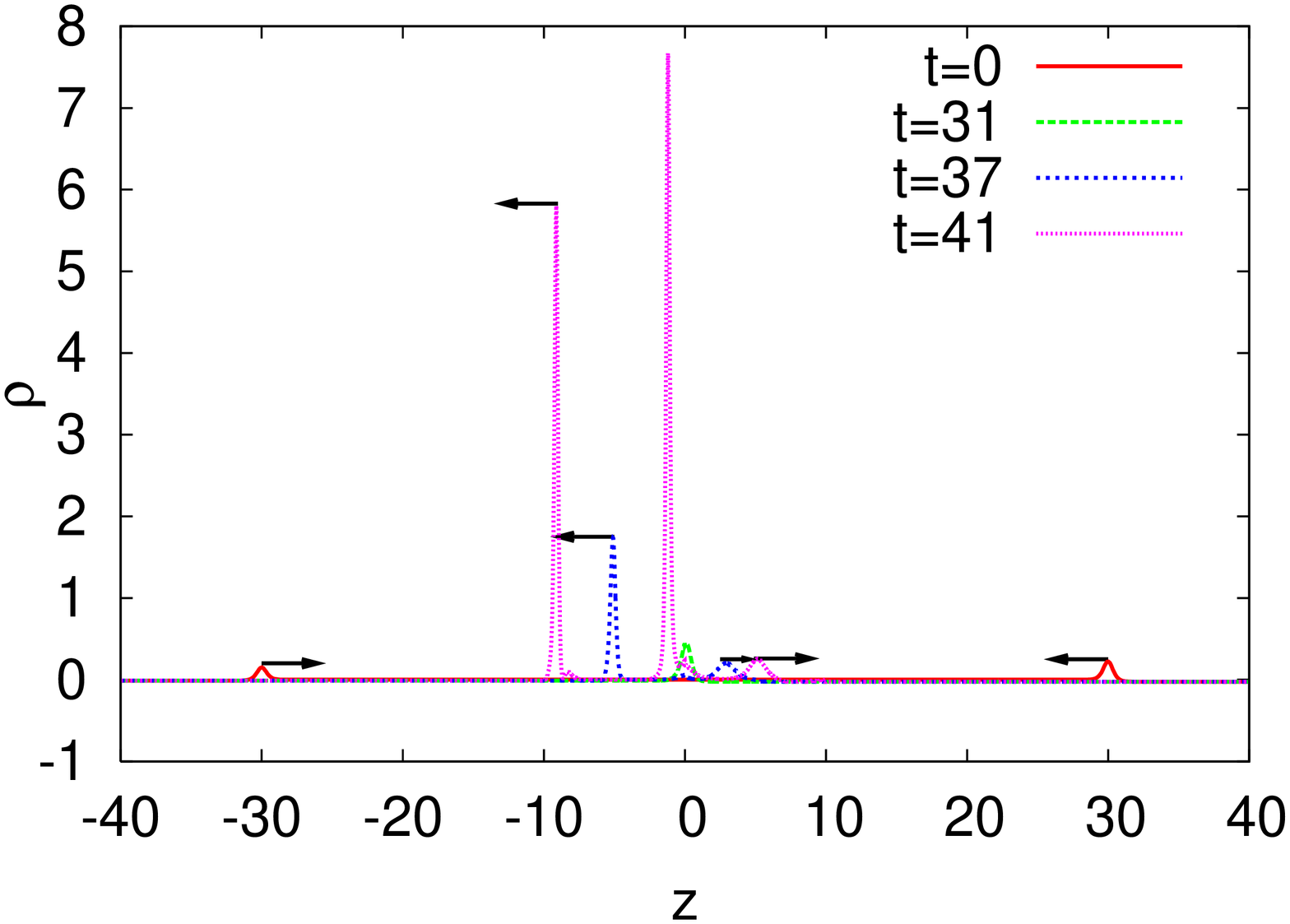}
\label{fig:rho vs z:C}
}
\vspace{-0.4cm}
\caption{
Energy density $\rho$ on $t=\mathrm{const.}$ surfaces:
(a) collision of two identical walls at the center, 
(b) collision of two walls with different thickness, 
and  
(c) two walls with different amplitude. 
The sharp peaks represent the domain walls, and the arrows show the directions of a wall's velocity. 
For (b) and (c), all the unspecified parameters are the same as those in (a).
\label{fig:rho vs z}
}
\end{figure*}
%%%%%%%%%%%%%%%%%%%%%%%%%%%%%%%%%%%%%%%%%%%%%%

%%%%%%%%%%%%%%%%%%%%%%%%%%%%%%%%%%%%%%%%%%%%%%
\begin{figure*}[bt]
\vspace{-0.3cm}
\includegraphics[width=4cm,angle=-90]{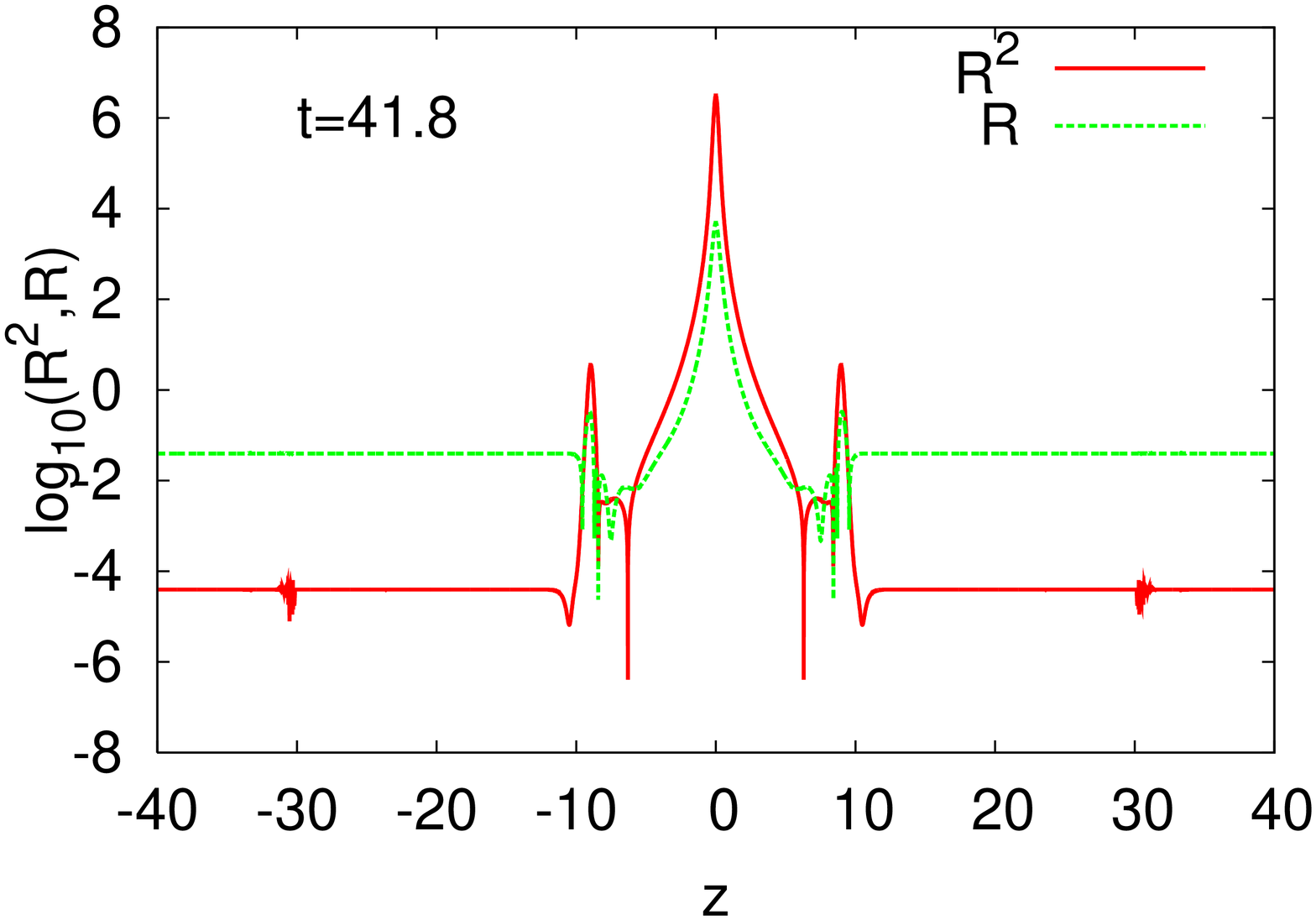}
\includegraphics[width=4cm,angle=-90]{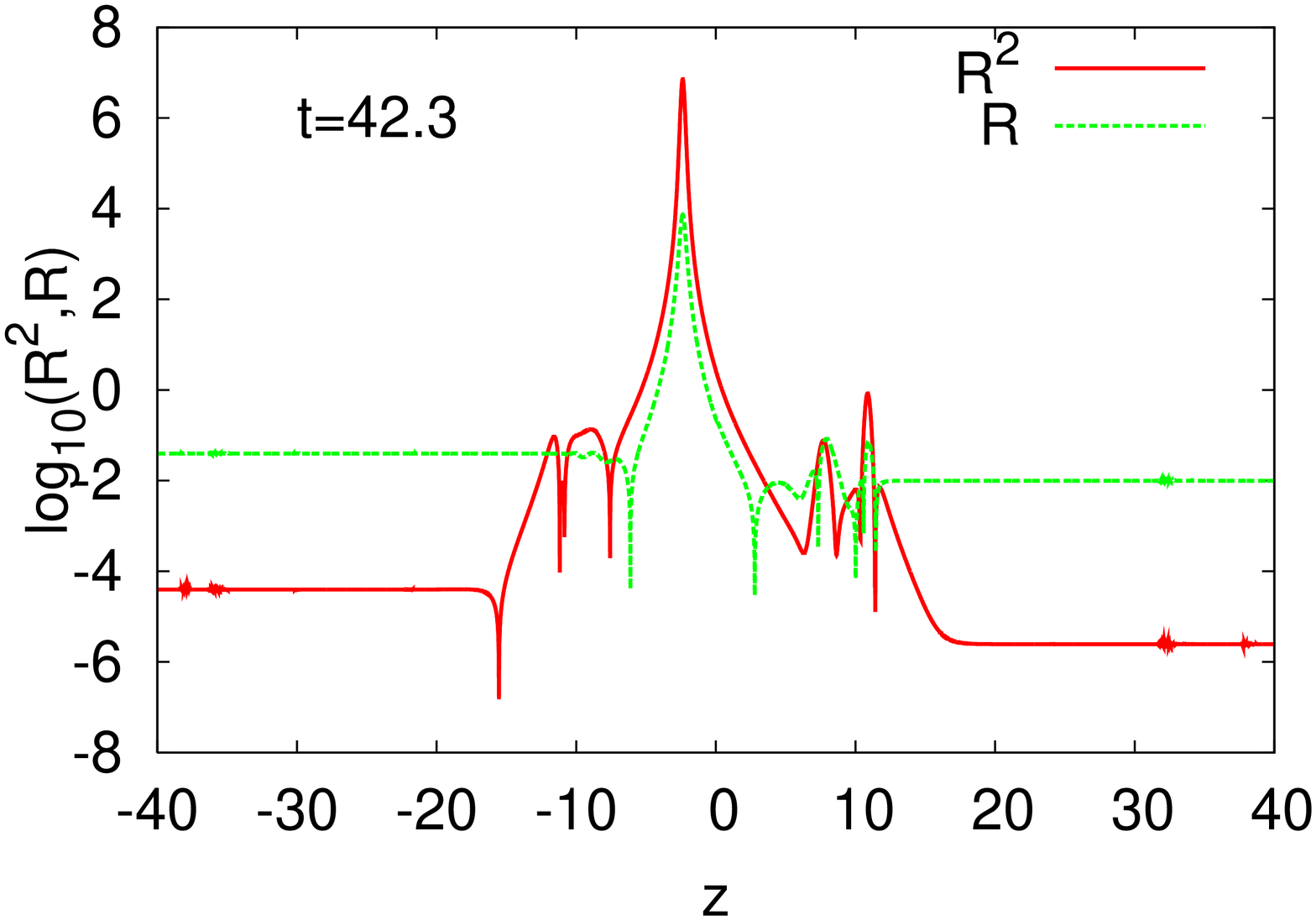}
\includegraphics[width=4cm,angle=-90]{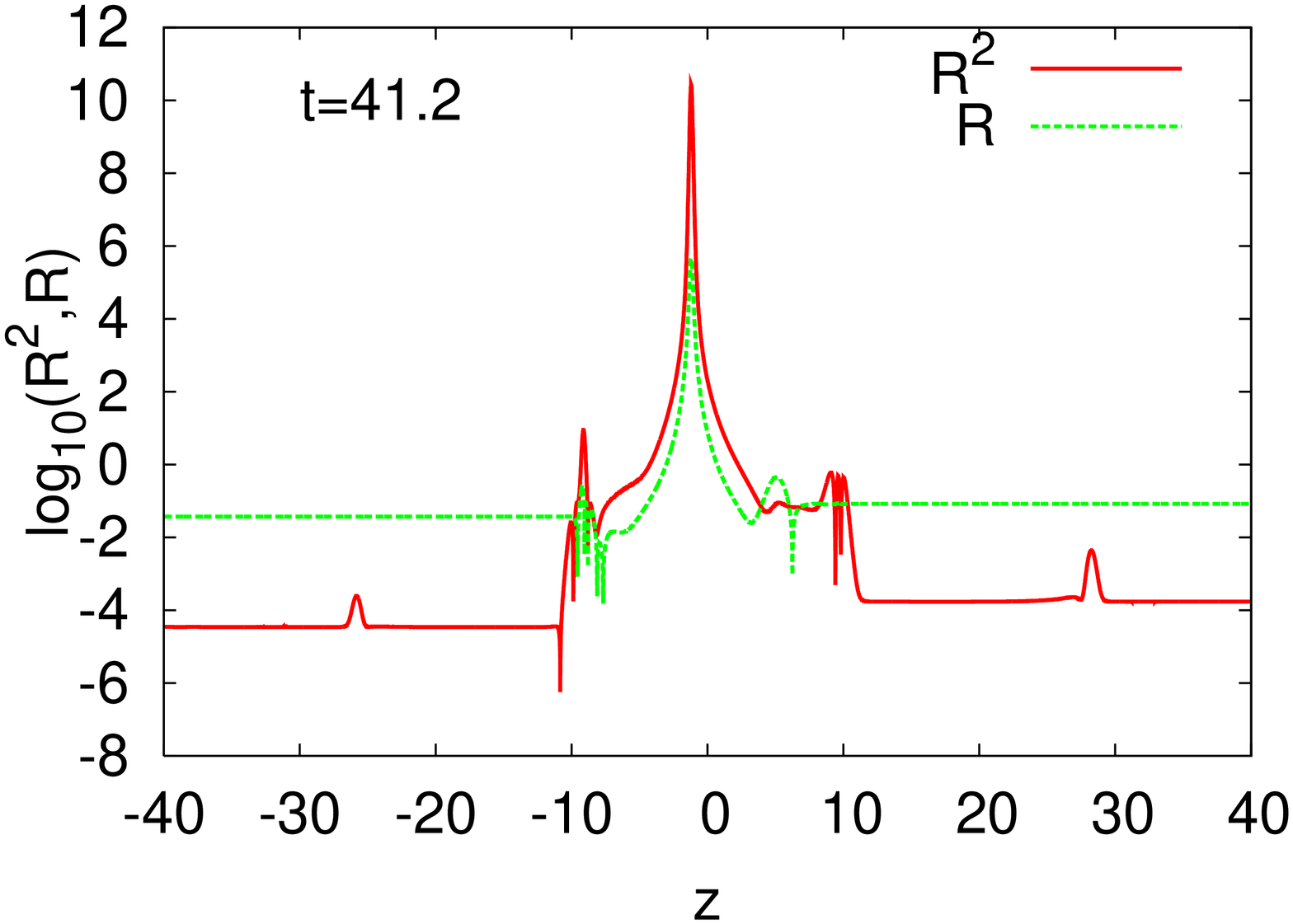}
\caption{
\label{fig:curvature}
$R$ and $R^{abcd}R_{abcd}$ at $t=$const. surface, corresponding to the simulations in Fig.~\ref{fig:rho vs z}. 
}
\end{figure*}
%%%%%%%%%%%%%%%%%%%%%%%%%%%%%%%%%%%%%%%%%%%%%%

%%%%%%%%%%%%%%%%%%%%%%%%%%%%%%%%
%%%%%%%%%%%%%  Fig.2   %%%%%%%%%
%%%%%%%%%%%%%%%%%%%%%%%%%%%%%%%%
\begin{figure*}[t]
\begin{center}
%%-------------------------
\subfigure[ 
]{
\includegraphics[width=4.1cm,angle=-90]{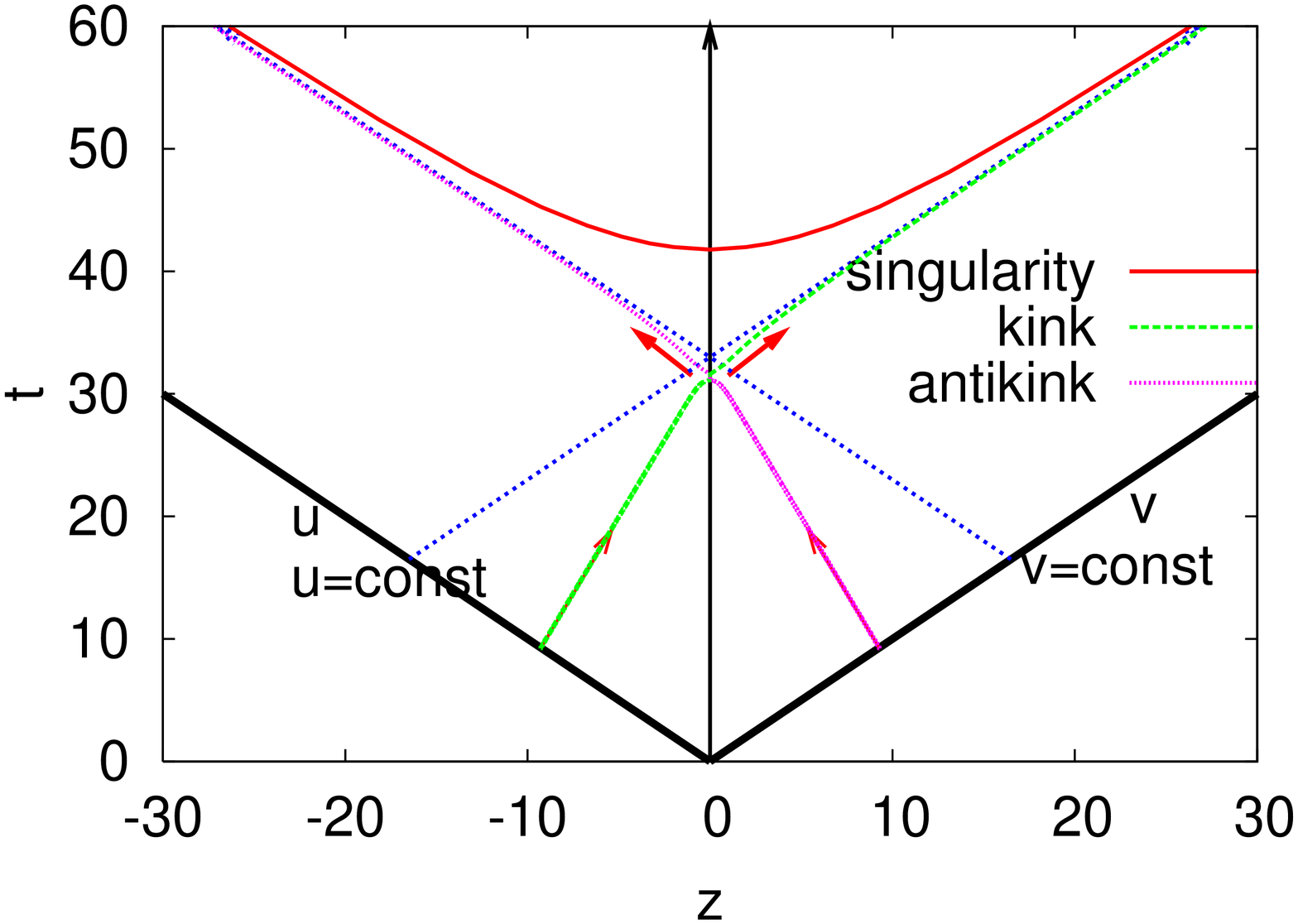}
\label{fig:uv}
}
\subfigure[ ]{
\hspace{-0.7cm}
\includegraphics[width=4.1cm,angle=-90]{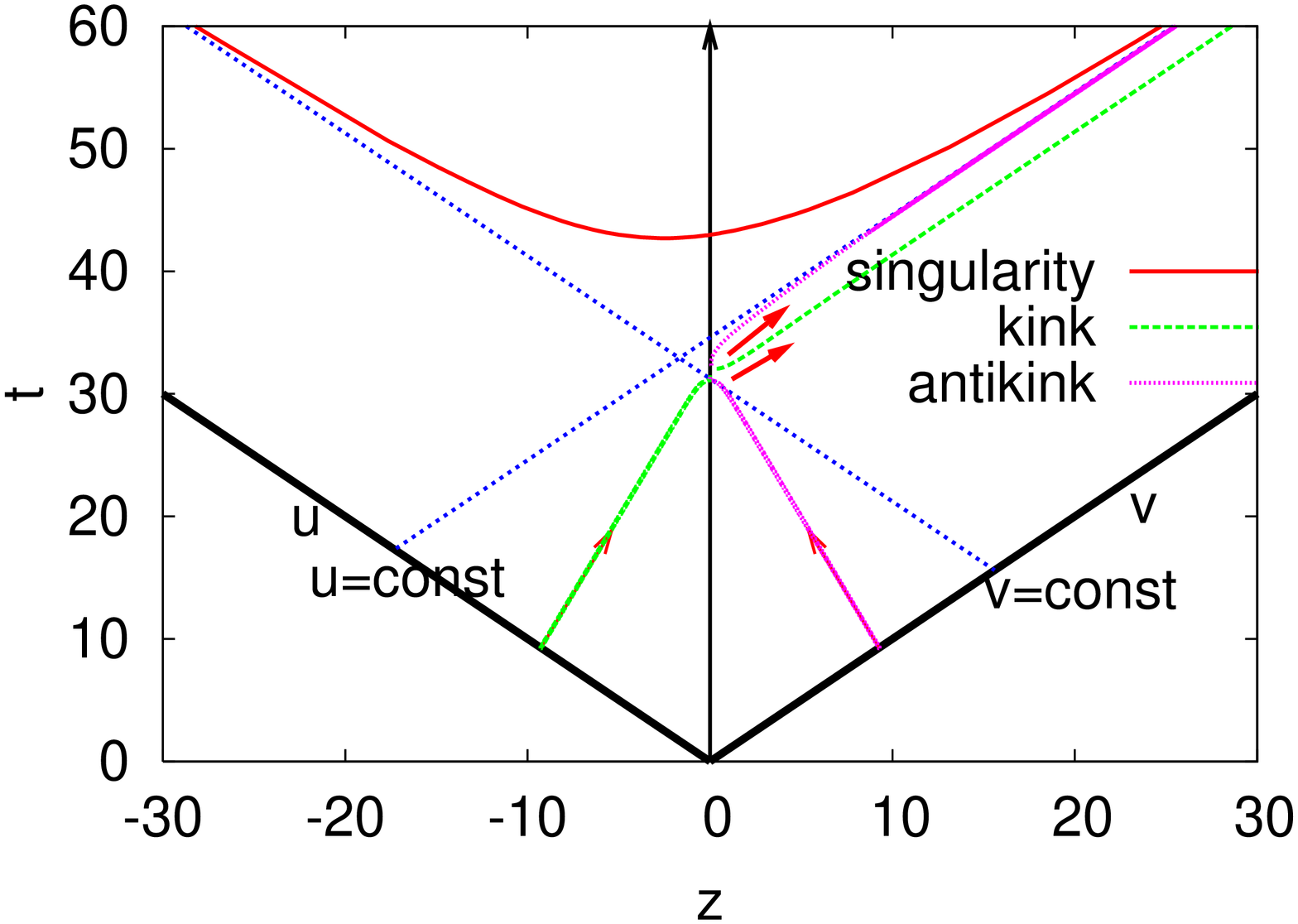}
\label{fig:uv2}
}
\subfigure[ 
]{
\hspace{-0.7cm}
\includegraphics[width=4.1cm,angle=-90]{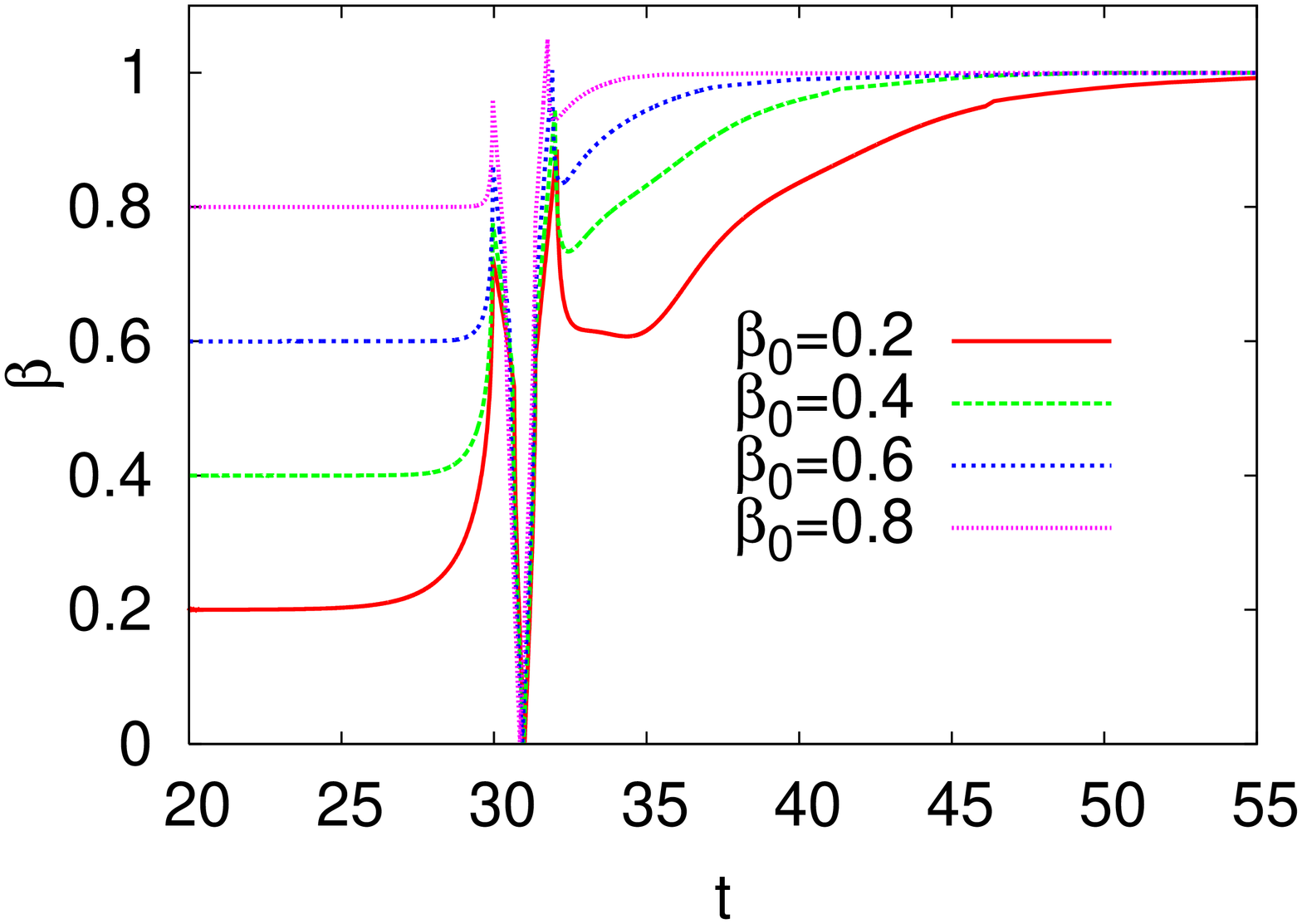}
\label{fig:wall velocity}
}
%%-------------------------
\vspace{-0.7cm}
\end{center}
\caption{
(a) and (b) show the numerical results in the null coordinates, corresponding to Fig.~\ref{fig:rho vs z:A} and~\ref{fig:rho vs z:B}, respectively. 
Dotted lines with arrows describe the orbit of walls. 
Thick lines are $u,v$-axes, and dotted straight lines show $u,v=$const., to which the spacelike singularity asymptotes. 
The position of a wall is defined by its maximum energy density. 
(c) Time variation of the ({kink}) wall's speed.
Irrespective of the initial speed, the final speed after the collision goes to unity.
}
\label{fig:uv-coordinate}
\end{figure*}
%%%%%%%%%%%%%%%%%%%%%%%%%%%%%%%%%%%%%

The system we intend to study consists of two domain walls that are initially located far away from each other. 
The initial data for such a configuration is constructed by superposing domain wall solutions in an appropriate manner.  
As a model of a domain wall, we consider a gravitationally interacting scalar field of the Lagrangian 
$
\mathcal{L} =
 \frac{R}{2{\kappa^2}} - \frac{1}{2} \nabla^a\phi\nabla_a\phi - V.
$
The following solution of 5D Einstein equations represent a single domain wall ({\it{Model~I}}), which has a spatially flat direction in three dimensions $d{\vec{x}_3}^2$. 
\begin{eqnarray}
  ds^2 &=&  e^{2U} (-dt^2 + d{\vec{x}_3}^2) + dr^2, 
\label{eq:model Arai}
\\
 U &=& - \frac{{L}^2}{3}  
    \Bigl[ \log\left[ \cosh\left( \frac{2(r-r_0)}{\delta} \right)  \right]
     +  \frac{2a (r-r_0)}{\delta} \Bigr], 
\nonumber
\cr
 \phi &=& \frac{2 {L}}{{\kappa}} 
    \left\{
  \tan^{-1} \left[ \tanh\left(  \frac{r-r_0}{\delta} \right)  \right]
   - \frac{\pi}{4} 
 \right\},
 \nonumber
\cr
 V(\phi) &=&  \frac{2{L}^2}{3 {\kappa^2} \delta^2}  
    \left[
      \omega_0 
    + \omega_1  \cos\left(  \frac{{\kappa} \phi}{{L}}  \right) 
    + \omega_2  \cos^2 \left(\frac{{\kappa} \phi}{{L}}  \right) 
    \right] ,
\nonumber
\end{eqnarray}
where 
$
\omega_0 = 3-4a^2 {L^2},~
\omega_1  = - 8 a {L^2},~ 
\omega_2= - 3- 4 {L^2}.
$
There are three unfixed parameters, i.e., wall thickness $\delta$, amplitude ${L}$ of scalar field, and the position $r_0$ of the kink's core. We will hereafter take $\kappa^2=1$. 
We will call this domain wall solution the {\it{kink}} solution (for $a<0$).  The {\it{anti-kink}} solution is defined by the reflecting $r$-coordinate in the above solution.

In the limit of $r\to \pm \infty$, the scalar field asymptotes constants, and the scalar potential plays the role of the cosmological constant $\Lambda= {\kappa^2} V$ in the limit
\begin{eqnarray}
   \Lambda  = - \frac{8 {L^4}}{3\delta^2} 
             \Bigl[ 1 + a ~ \mathrm{sign}(r) \Bigr]^2_{r \to \pm \infty}. 
\label{eq:cosmological constant}
\end{eqnarray}
The domain wall for $|a| < 1$ gives a warp factor decreasing for both infinities of an extra dimension, and the cases of $|a| = 1$ become the wall solutions interpolating between AdS and flat Minkowski vacua. 
For $|a|>1$, the warp factor decreases in one direction, and increases in the other.
This domain wall can be embedded into the five-dimensional supergravity coupled with hypermultiplets as an exact BPS domain wall \cite{Arai:2002ph}.
After integrating out irrelevant fields with canonical normalization, the above solution is found to be identical to the exact BPS solution in \cite{Arai:2002ph}.

We shall restrict our analysis to collisions along a $r$-direction, preserving the symmetry along the homogeneous $\vec{x}_3$-directions.
Even with this simplification, such a setup is of relevance in a number of physical situations. 
The initial data for such a collision can be obtained as follows.  
First of all, we introduce a new coordinate $z$ by $z=\int dr e^{-U}$ and work on the conformal gauge, 
\begin{align}
ds^2 = e^{2A(t,z)}(-dt^2+dz^2)
         + e^{2B(t,z)}d\Vec{x}^2 .
\end{align}
Then the above single static wall is boosted along the fifth direction $z$, and we obtain a wall moving with constant velocity $\beta$ \cite{Takamizu:2006gm}.

To discuss collisions of two moving domain walls, we set a kink solution at $z=-z_0$ and an anti-kink solution at $z=z_0$, which are separated by a large distance and approaching each other with the same (or different) speed $\beta$. Such superposition and matching of the metric and scalar field at the center is possible for $|a|=1$, and sufficiently smooth initial data that satisfies the constraint equations at the initial time can be obtained, as long as the spatial separation between the two walls is much larger than the thickness of walls.  Therefore, we take $|a|=1$ throughout this paper. 
Obviously, we set $A=B$ and velocity $\dot{A}=\dot{B}$ at  {the outset}, and the initial values of $\dot{\phi}$ and $\dot{A}$ are given by the above construction. 
During the evolutions, the Neumann boundary conditions are imposed at the outer boundaries.
The asymptotics of the scalar field is {$\phi 
\propto (\gamma|z| /\sqrt{6/|\Lambda|} +1)^{-3/2 L^2}$}, 
and the metric behaves {$e^{A} \propto (\gamma |z|/\sqrt{6/|\Lambda|}+1)^{-1}$} as 
$|z|\to \infty$, where $\gamma=1/\sqrt{1-\beta^2}$ is the Lorentz factor.
The kink and anti-kink solutions are characterized by their own width and amplitude, $(\delta_K, {L}_K)$ and $(\delta_A, {L}_A)$, respectively.
Therefore, we have three types of unfixed parameters for the initial setup; $\delta$, ${L}$ and $\beta$.

Using a fourth-order accurate finite difference code, we have solved the system numerically and evaluated the constraints at each time step for various families of initial data. 
The overall picture does not depend on a specific choice of the parameters. 
Some examples of numerical results are reported in Fig.~\ref{fig:rho vs z}, in which 
 the time evolutions of energy density 
\begin{eqnarray}
 \rho = \frac{ e^{-2A}}{2}
  \Bigl[  (\partial_t\phi)^2 + (\partial_z \phi )^2 
  \Bigr] 
 + V
\end{eqnarray}
in $\{t,z\}$-flame are shown. 
In all these cases, the two walls with initial velocity $\beta_0=0.4$ collide at $z=0$ and $t \approx 31$. 
Fig.~\ref{fig:rho vs z:A} describes the symmetric collision of two identical walls. 
In this case, the walls {\it{pass through}} one another so that the initial kink solution at $z<0$ goes to $z>0$. (The kink and anti-kink solutions are distinguished by their relative field values at the center and infinity $z=\pm\infty$.)
The energy density of wall, i.e. the wall's tension, increases during this process. 
This would be caused by the fact that the induced universe on the walls are contracting during the process, with $\dot{B}<0$. After the collision, a sharp peak of density appears at the collision point $z=0$, and it implies an emergence of singularity.
In fact, the curvature diverges rapidly at the point, whereas the curvature on the wall remains finite and small at the moment (Fig. \ref{fig:curvature}). 
Here our criterion of curvature singularity is that the Kretschmann scalar exceeds $R^{abcd}R_{abcd} > 10^6$. 
At the time $t\approx 42$ of singularity formation, the energy density localized at $z=0$ is 1.2 times bigger than those on the walls, and a portion of energy is stored in this small region, which will be inside an event horizon, as we see below.

This basic picture of collision holds for other cases. 
For the asymmetric collisions, such as two walls with different width, 
amplitude, and/or speeds, the emergence of singularity is still a generic feature. Figs.~\ref{fig:rho vs z:B} and~\ref{fig:rho vs z:C} show examples of collisions in which different thickness or amplitude of scalar field are taken for the two walls, without changing other parameters. 
Among these cases, Fig.~\ref{fig:rho vs z:B} shows that one of the walls recoils at the collision, due to the larger momentum of one of the walls: 
the initial kink solution at $z=-30$ in Fig.~\ref{fig:rho vs z:B} goes to $z>0$ after the collision, while the anti-kink at $z=+30$ bounds back. 
Interestingly, in these asymmetric collisions, the curvature singularities appear off the collision point. 
For Figs.~\ref{fig:rho vs z:B} and~\ref{fig:rho vs z:C}, they are at $z=-2.4$ and $z=-1.2$, respectively.

For the wide range of initial parameters, the emergence of singularity is the generic consequence. However, as expected and discussed below, the singularity does not appear for ${L} \ll 1$ and/or $\beta_0 \ll 1$ for fixed $\delta$. In such ``non-relativistic'' cases, the two walls just pass through, and the final configurations of fields are well described by the boosted walls, as we applied for the initial configurations.

\vspace{-0.5cm}
%%--------------------------------------------------------
\subsection{Horizon formation}
%%--------------------------------------------------------

The next task at hand is to confirm the nature of singularity. 
In numerical investigations of singularity formation and global structure of the spacetime, null coordinates are useful to prevent the singularity from corrupting the rest of the spacetime. 
In these coordinates, horizons are not particularly special and we can follow the collision all the way to the singularity even when a horizon appears through a collision. 
We evolve the colliding walls in the double-null coordinates (e.g., \cite{Frolov:2004rz,Frolov:2003dk,Hamade:1995ce}), 
\begin{equation}
ds^2 
     = -2 e^{2A}dudv+e^{2B}d\Vec{x}^2 \,,
\end{equation}
where $\sqrt{2}u=(t-z)$, $\sqrt{2}v=(t+z)$.
In this gauge, the Einstein equations and the dynamical equation for a 
scalar field are split into three dynamical 
and two constraint equations.
%
%%%\begin{eqnarray}
% &&
%  2\partial_u\partial_v A
% -6\partial_u B\partial_v B 
% +{\kappa^2}
% \Bigl(\partial_u \phi\partial_v \phi+{1\over 3}e^{2A}V \Bigr)=0\,,
% \nonumber
% \\
% && \partial_u\partial_v B+3\partial_u B\partial_v B -{1\over 3} {\kappa^2}
% e^{2A}V =0\,,
% \nonumber
% \\
% && 
% 2\partial_u\partial_v \phi 
% +3\partial_u \phi\partial_v B 
% +3\partial_u B\partial_v \phi
% +e^{2A}V' =0\,,
% \label{eq:dynamical eqs}
% \end{eqnarray}
%

Let us first focus on the symmetric collision in Fig.~\ref{fig:rho vs z:A}. The corresponding evolution in the null coordinates is described in Fig.~\ref{fig:uv-coordinate}.  
It shows that the curvature singularity is spacelike, approaching $u$ or $v=\mathrm{const.}$ lines at late times, which corresponds to the event horizon.
This result is very generic, and we have observed similar results for the wide range of initial data (velocity, etc.). 
This system has homogeneous 3-spatial directions, and so the horizon also extends in these directions. This means that a black brane is produced by the collision of walls, so that this collision provides the dynamical mechanism of generating black branes in higher dimensions.

\begin{figure}[t]
\begin{center}
\includegraphics[width=4.2cm]{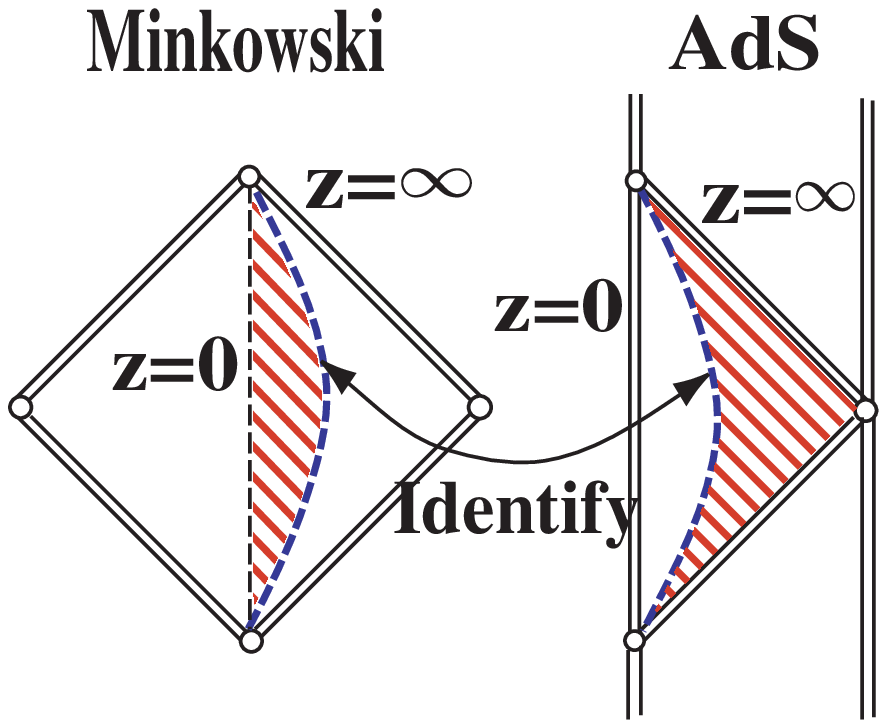}
\hspace{0.4cm}
\includegraphics[width=3.3cm]{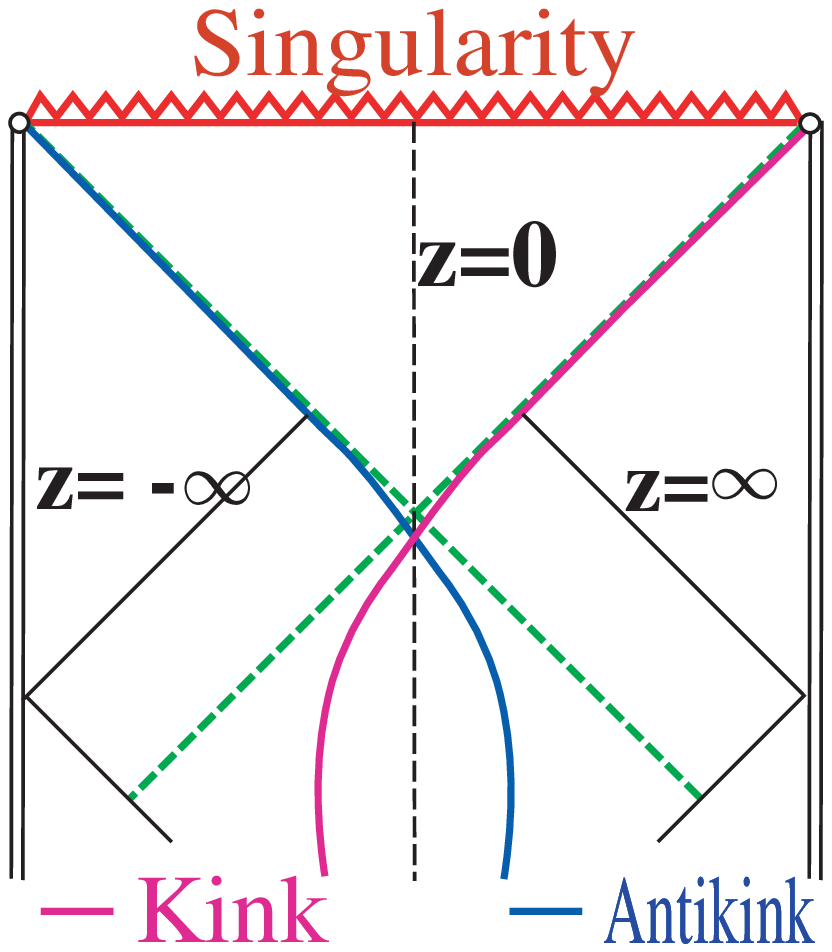}
\end{center} 
\vspace{-0.5cm}
\caption{
Schematic conformal diagrams for a single domain wall that asymptotes to AdS (Minkowski) as $z\to \infty$ ($z\to 0$) {\it{(Left)}} and for  
colliding walls, producing a black brane {\it{(Right)}}.
\label{fig:conformal diagram}
}
\end{figure}

Another interesting feature is that after the collision the walls are trapped around the surface of the horizon. 
 The final speeds of walls asymptote to the speed of light [Fig.~\ref{fig:wall velocity}], irrespective of the initial velocity. 
The bulk outside the two walls is not exactly AdS, but asymptotes to it. 
Because of this  {behavior}, the walls are pulled outside, accelerating in the directions.  A schematic picture of a conformal diagram is given in Fig.~\ref{fig:conformal diagram}

Another example corresponding to to the asymmetric collision in Fig.~\ref{fig:rho vs z:B} is shown in Fig.~\ref{fig:uv2}. 
Even in this asymmetric collision, the event horizon forms from the point where the spacelike singularity appears. 
An interesting difference is that the kink wall escapes from the horizon, and only the antikink wall is trapped  {nearby}.

%%%%%%%%%%%%%%%%%%%%%%%%%%%%%
%%%%%%%%%%%%%%%%%%%%%%%%%%%%%

\vspace{-0.5cm}
%%--------------------------------------------------------
\subsection{Model II}
%%--------------------------------------------------------

The initial data discussed so far is based on the single BPS domain wall. 
There is another simple model in which collisions of walls can be tested. 
It is the model used in the previous work \cite{Takamizu:2006gm},
\begin{eqnarray}
&& U = - \frac{2 {L^2} }{3}\left( \log[ \cosh(r/\delta) ]
      + \frac{\tanh^2(r/\delta)}{4}
       {-} \frac{r}{\delta}
        \right), 
\cr
&& \phi =  \frac{{\sqrt{3}}~ {L}}{{\kappa}} 
          \tanh\left( \frac{r}{\delta} \right) ,
\cr
&& V= \frac{9{L}^4}{4\delta^2{\kappa^4}}
       \left[ 2 \left(\frac{\partial W}{ \partial \phi} \right)^2 
         - \frac{8{\kappa^2}}{3}   W^2
       \right],
\label{eq:Model TM}
\end{eqnarray}
where
$W =   {-}\frac{1}{9\sqrt{3}} \left( \frac{{\kappa}}{L} \phi \right)^3
      {+}\frac{ {\kappa} }{\sqrt{3} L }\phi  
     -\frac{2}{3}
$.
The basic property of the wall is quite similar to the wall in the previous sections;
the bulk in  {$r>0$} asymptotes to the Minkowski spacetimes, while the spacetime in  {$r<0$} asymptotes to the AdS, recovering (\ref{eq:cosmological constant}). This single domain wall solution is found simply by extending the four-dimensional solution in \cite{Eto:2003bn}.

It is interesting to study various aspects of the wall collisions in this model and compare them with the previous model.
We have performed many simulations and confirmed that all phenomena observed in the previous sections, such as the singularity and horizon formation hold with qualitatively similar behaviors. 
A basic exception is that in this model the walls {\it{bounce back}} after the collision, contrary to the case in Model I. 
Thus the causal structures of such a collision look like Fig.~\ref{fig:uv}, but kink and anti-kink profiles are exchanged after a collision. 
This difference comes from the nonlinear interaction through the dynamics of collision.

\begin{figure}[t]
\begin{center}
\includegraphics[width=4.2cm,angle=-90]{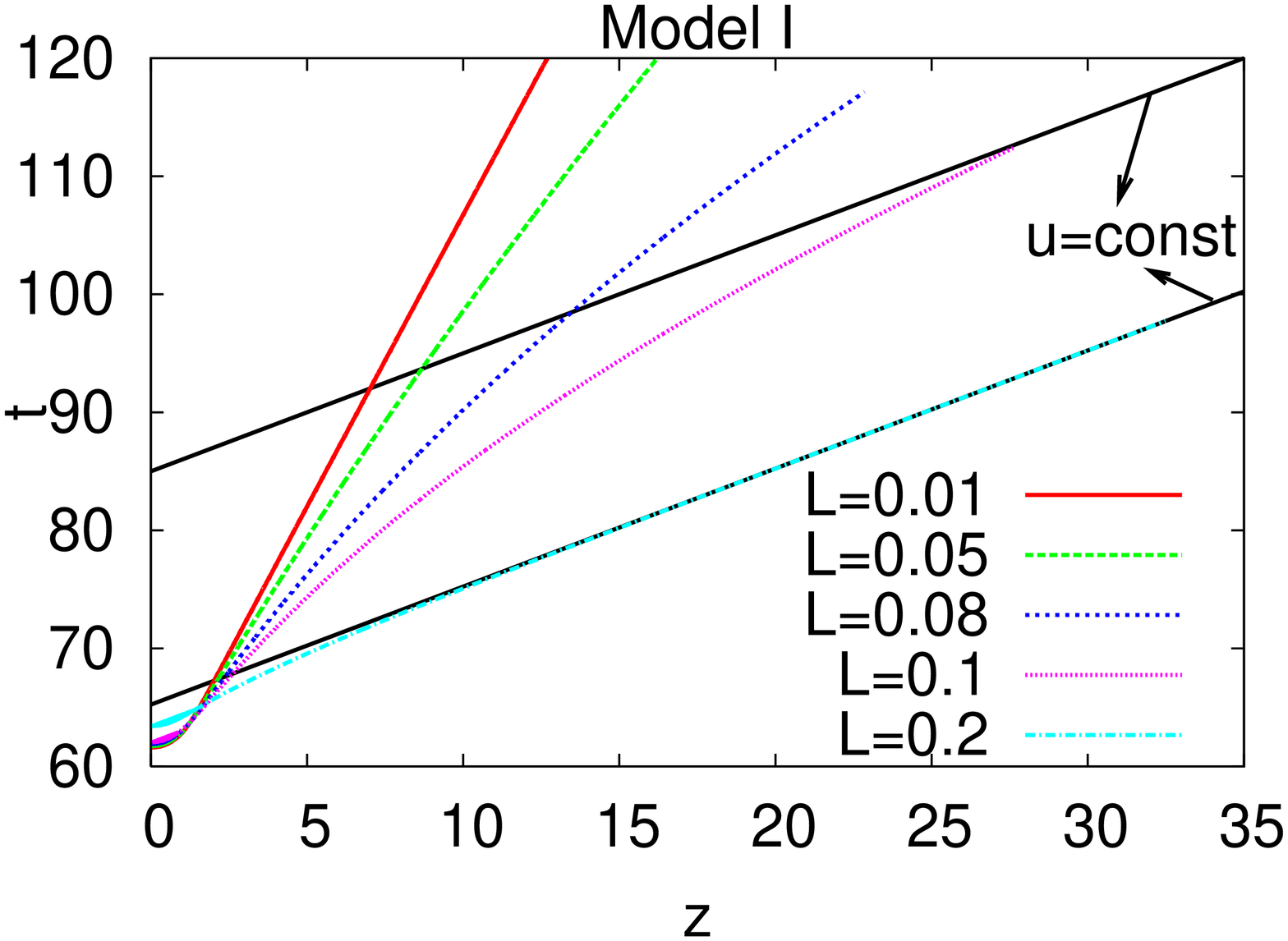}
 \hspace{-0.6cm}
\includegraphics[width=4.2cm,angle=-90]{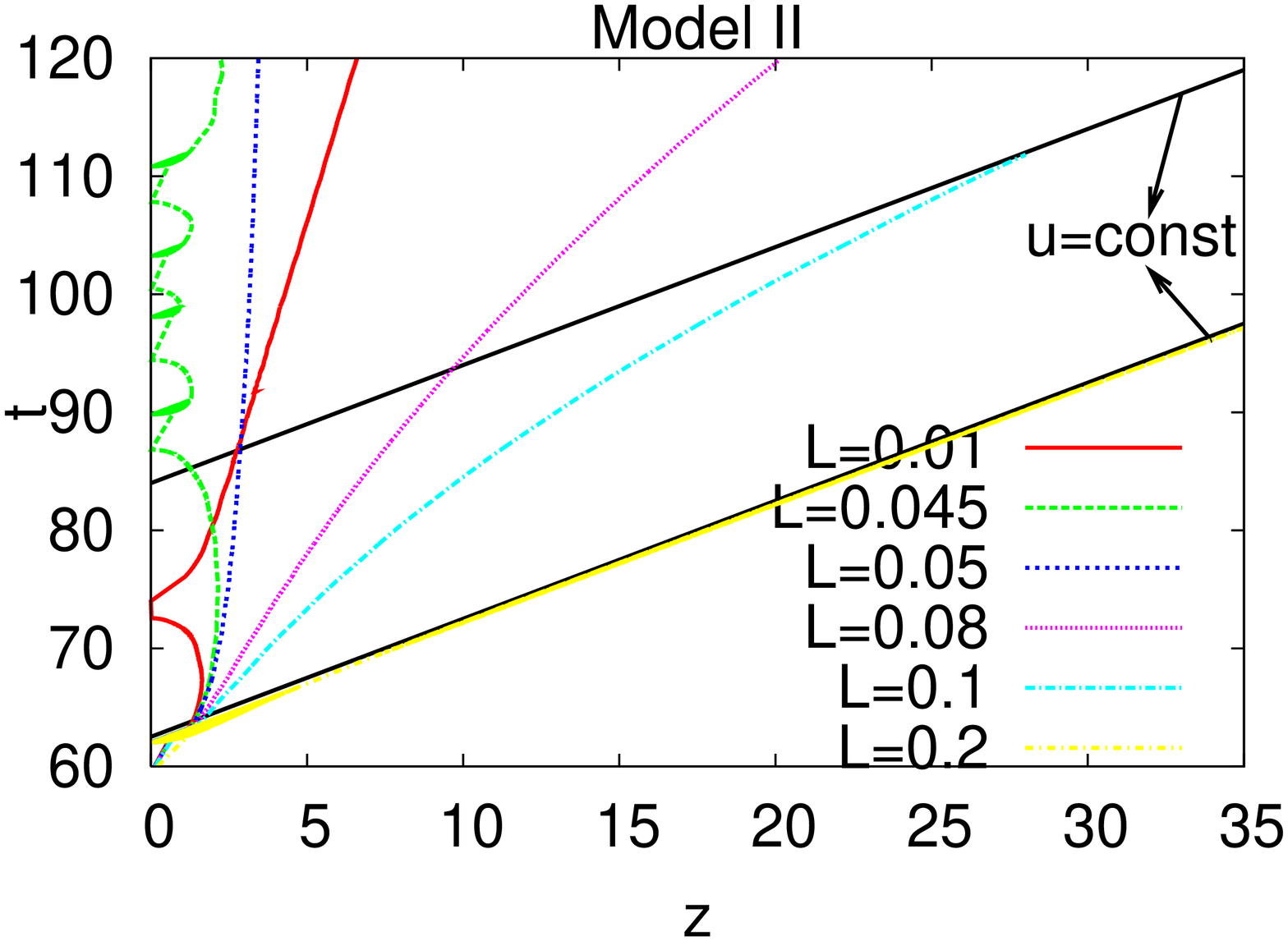}
\end{center}
\vspace{-0.4cm}
\caption{
The orbits of a wall after the symmetric collisions at $t \approx 60$ ($\beta_0=0.2$ and $\delta=1$) are described  {for the two models (Model I, II)}, to show the dependence on ${L}$. 
Solid straight (black) lines represent $u=$const. lines along the spacelike singularities for ${L}=0.1, 0.2$. 
\label{fig:Gamma dependence}
}
\end{figure}

Such details of model dependence become more significant for weak field cases in which no singularity appears. 
In Fig.~\ref{fig:Gamma dependence}, we compare the difference of the two models by showing the orbits of a wall after the symmetric collisions for various 
values of ${L}$. 
For Model I, the spacelike singularity appears for ${L} \gtrsim 0.1$, and the walls asymptote to the null lines, as discussed above.  
On the other hand, for ${L} \lesssim 0.1$, the velocity of the wall becomes   timelike with constant speed after the collisions, and no singularity appears. 
In fact, the final configurations of scalar fields are well approximated by superposing boosted walls, so that the two walls just pass through one another in these cases. 
Note that if the initial velocity is increased the horizon appears even for smaller ${L}$.

For Model II, multiple collisions take place for ${L} \ll 1$ (Fig.~\ref{fig:Gamma dependence}).
For ${L}=0.01$, the collision takes place {two times}, and then the wall 
bounces back with constant velocity. 
This behavior is compatible with and typical in the non-gravitating system of 
the previous study \cite{Takamizu:2006gm}. 
As ${L}$ increases, the two walls gravitate toward one another and multiple bounces take place (e.g., ${L}=0.045$). 
The marginal value of ${L}$ is ${L}=0.05$ in Fig.~\ref{fig:Gamma dependence}, in which the gravitational attractive force and the repulsive force due to outer AdS region are in balance. 
Therefore, a quasi-static configuration of two walls is realized after the collision.

%%%%%%%%%%%%%%%%%%%%%%%%%%%%%
%%%%%%%%%%%%%%%%%%%%%%%%%%%%%

\vspace{-0.5cm}
%%--------------------------------------------------------
\section{Summary and Discussion}
%%--------------------------------------------------------

We have considered a system of colliding domain walls, whose initial data is set up based on a single BPS domain wall, and found that the role of gravity at the collision is significant in that it can drastically change the picture of ``silent" collisions without self-gravity, as observed also in colliding (impulsive) gravitational waves \cite{Griffiths}.
The main result of our study is that horizon formation is a generic phenomena in the collision of walls. 
In the non-relativistic cases, such as ${L} \ll1$ and/or $\beta \ll 1$,  
``silent" collisions without singularity and horizon are realized, but such cases are very limited and unlikely in an early universe of braneworlds.

The local interaction and dynamics at the collision depends on which model we are looking at. 
For Model I, the domain walls can pass through one another, while in Model II they bounce at the collision and go back. 
However, the basic feature of horizon formation does not change in these models, and we have concluded that the horizon formation (and singularity behind it) is a generic consequence of kink-anti-kink collisions. 
The horizon has three homogeneous spatial directions, so that a black three-brane is produced by such a collision. 
The bulk outside the two walls is approximated by the AdS, and then we look at the created black hole as it sits on the AdS. 
In the pure AdS, a possible black hole is a topological black hole, which has a flat 3-dimensional hypersurface with vanishing curvature. The end state of the present scenario would be this type of black hole (Fig.~\ref{fig:conformal diagram}), although the field fills the bulk outside the horizon. 
Here, a further interesting possibility comes from the fact that a spatially homogeneous horizon suffers from Gregory-Laflamme instability in general. The end state of this instability has not been clearly understood so far, and the horizon may break up, resulting in multiple black holes that are stuck on the walls \cite{Kudoh:2003xz}. 
Furthermore, during this process, a good deal of energy will be radiated away by gravitational waves, and they remain as primordial gravitational wave backgrounds. 
Thus this possibility provides a new way of producing primordial black holes and gravitational waves in an early universe with higher dimensional bulk filled by walls/branes. This scenario is analogous to the bubble collisions in a four-dimensional inflationary universe~\cite{Hawking:1982ga}.

There are several interesting directions, which may be pursued on the basis of these results and/or by relaxing several conditions. 
One of such issue is a study of the effects and roles of other fields in supergravity,  {a subject sets aside} in our analysis. 
Other fields contained in the hypermultiplets will be excited (or thermalized) during the collisions, and  {analyzing them should} provide many cosmological insights into braneworld cosmologies \cite{Gibbons:2006ge}. 
Furthermore, there is one most important question left to answer: 
 {To what extent} is the production of black holes/branes generic in a class of more generic theory and context, such as collisions of different types of walls with an arbitrary incident angle. We  {foresee} no major obstacle in anticipating that horizon formation  {would} be suppressed.

\vspace{-1cm}
%%================================================================
\acknowledgments
We would like to thank G.T.~Horowitz and S. Mukohyama for their comments and discussions. The work is supported by JSPS, Japan-U.K. Research Cooperative 
Program and the 21COE Program at Waseda University.
%%================================================== 

%%%%%%%%%%%%%%%%%%%%%%%%%%%%%%%%%%%%%%%%%%%%%%%%%%%%%%%%%%%%%%%
%%%%%%%%%%%%%%%          REFERENCES          %%%%%%%%%%%%%%%%%%
%%%%%%%%%%%%%%%%%%%%%%%%%%%%%%%%%%%%%%%%%%%%%%%%%%%%%%%%%%%%%%%
%%%%----------------------------------------------------
% \bibliographystyle{plain}
 \bibliographystyle{apsrev} 
% \bibliographystyle{unsrt} %% plain, unsrt, alpha, abbrv, acm, apalike
% \bibliography{referenceBH_200701}

\begin{thebibliography}{21}
\expandafter\ifx\csname natexlab\endcsname\relax\def\natexlab#1{#1}\fi
\expandafter\ifx\csname bibnamefont\endcsname\relax
  \def\bibnamefont#1{#1}\fi
\expandafter\ifx\csname bibfnamefont\endcsname\relax
  \def\bibfnamefont#1{#1}\fi
\expandafter\ifx\csname citenamefont\endcsname\relax
  \def\citenamefont#1{#1}\fi
\expandafter\ifx\csname url\endcsname\relax
  \def\url#1{\texttt{#1}}\fi
\expandafter\ifx\csname urlprefix\endcsname\relax\def\urlprefix{URL }\fi
\providecommand{\bibinfo}[2]{#2}
\providecommand{\eprint}[2][]{\url{#2}}


%\cite{Khoury:2001wf}
\bibitem{Khoury:2001wf}
  J.~Khoury, B.~A.~Ovrut, P.~J.~Steinhardt and N.~Turok,
  %``The ekpyrotic universe: Colliding branes and the origin of the hot big bang,''
  Phys.\ Rev.\  D {\bf 64}, 123522 (2001);
  %%[arXiv:hep-th/0103239].
%%==============================--
%\cite{Steinhardt:2002ih}
%\bibitem{Steinhardt:2002ih}
  P.~J.~Steinhardt and N.~Turok,
  %``A cyclic model of the universe,''
  Science {\bf 296}, 1436 (2002).

%%------------
%\cite{Takamizu:2006gm}
\bibitem{Takamizu:2006gm}
  Y.~Takamizu and K.~Maeda,
  %``Collision of domain walls in asymptotically anti de Sitter spacetime,''
  Phys.\ Rev.\  D {\bf 73}, 103508 (2006);
  %% [arXiv:hep-th/0603076].
%% ===================
%% \bibitem{Takamizu:2004rq}
    ibid.
%%   Y.~Takamizu and K.~Maeda,
  %``Collision of domain walls and reheating of the brane universe,''
  Phys.\ Rev.\  D {\bf 70}, 123514 (2004)
%%------------


%%------------ Domain wall dynamics -----------
%\cite{Shinkai:1993vk}
\bibitem{Shinkai:1993vk}
  H.~Shinkai and K.~Maeda,
  %``Can gravitational waves prevent inflation?,''
  Phys.\ Rev.\  D {\bf 48}, 3910 (1993);
  %% [arXiv:gr-qc/9305014].
%%===========================
%\cite{Flachi:2006ev}
 % \bibitem{Flachi:2006ev}
  A.~Flachi, O.~Pujolas, M.~Sasaki and T.~Tanaka,
  %``Critical escape velocity of black holes from branes,''
  Phys.\ Rev.\  D {\bf 74}, 045013 (2006);
  %% [arXiv:hep-th/0604139].
%%===========================
%\cite{Flachi:2005hi}
% \bibitem{Flachi:2005hi}
  A.~Flachi and T.~Tanaka,
  %``Escape of black holes from the brane,''
  Phys.\ Rev.\ Lett.\  {\bf 95}, 161302 (2005);
  %% [arXiv:hep-th/0506145].
%%===========================
 %% \bibitem{Frolov:2003mc}
  V.~P.~Frolov, M.~Snajdr and D.~Stojkovic,
  %``Interaction of a brane with a moving bulk black hole,''
  Phys.\ Rev.\  D {\bf 68}, 044002 (2003);
  %% [arXiv:gr-qc/0304083].
  %%CITATION = PHRVA,D68,044002;%%
%%===========================
%\bibitem{Morisawa:2002br}
  Y.~Morisawa, D.~Ida, A.~Ishibashi and K.~Nakao,
  %``Thick domain walls around a black hole,''
  Phys.\ Rev.\  D {\bf 67}, 025017 (2003);
  %% [arXiv:gr-qc/0209070].
%%===========================
%\cite{Martin:2004bj}
%%\bibitem{Martin:2004bj}
  J.~Martin, G.~N.~Felder, A.~V.~Frolov, L.~Kofman and M.~Peloso,
  %``BRANECODE: A program for simulations of braneworld dynamics,''
  Comput.\ Phys.\ Commun.\  {\bf 171}, 69 (2005);
  %% [arXiv:hep-ph/0404141].
%\cite{Antunes:2003kh}
%%===========================
%\bibitem{Antunes:2003kh}
  N.~D.~Antunes, E.~J.~Copeland, M.~Hindmarsh and A.~Lukas,
  %``Kink-boundary collisions in a two dimensional scalar field theory,''
  Phys.\ Rev.\  D {\bf 69}, 065016 (2004);
  ibid
  Phys.\ Rev.\  D {\bf 68}, 066005 (2003).
  %[arXiv:hep-th/0208219].
  %%CITATION = PHRVA,D68,066005;%%
%%-----------------------------------------------------


%%---------- Bubbles  ----------------
\bibitem{Blanco-Pillado:2003hq}
  J.~J.~Blanco-Pillado, M.~Bucher, S.~Ghassemi and F.~Glanois,
  %``When do colliding bubbles produce an expanding universe?,''
  Phys.~Rev.~D {\bf 69}, 103515 (2004);
  %% [arXiv:hep-th/0306151].
  %%------------%% 
  %%\bibitem{Copsey:2006qb}
  K.~Copsey,
  %``Bubbles unbound: Bubbles of nothing without Kaluza-Klein,''
  hep-th/0610058;
  %%CITATION = HEP-TH/0610058;%%
  %%------------%% 
  %%\bibitem{Horowitz:2002cx}
  G.~T.~Horowitz and K.~Maeda,
  %``Colliding Kaluza-Klein bubbles,''
  Class.\ Quant.\ Grav.\  {\bf 19}, 5543 (2002)
 %% [arXiv:hep-th/0207270].
  %%CITATION = CQGRD,19,5543;%%
%%-------------------------------------



\bibitem[{\citenamefont{Arai et~al.}(2003)\citenamefont{Arai, Fujita, Naganuma,
  and Sakai}}]{Arai:2002ph}
\bibinfo{author}{\bibfnamefont{M.}~\bibnamefont{Arai}},
  \bibinfo{author}{\bibfnamefont{S.}~\bibnamefont{Fujita}},
  \bibinfo{author}{\bibfnamefont{M.}~\bibnamefont{Naganuma}}, \bibnamefont{and}
  \bibinfo{author}{\bibfnamefont{N.}~\bibnamefont{Sakai}},
  \bibinfo{journal}{Phys. Lett.} \textbf{\bibinfo{volume}{B556}},
  \bibinfo{pages}{192} (\bibinfo{year}{2003}).
  %%, \eprint{hep-th/0212175}.


%%-------------------------------------
\bibitem[{\citenamefont{Frolov}(2004)}]{Frolov:2004rz}
\bibinfo{author}{\bibfnamefont{A.~V.} \bibnamefont{Frolov}},
  \bibinfo{journal}{Phys. Rev.} \textbf{\bibinfo{volume}{D70}},
  \bibinfo{pages}{104023} (\bibinfo{year}{2004});
%%, \eprint{hep-th/0409117}
%\bibitem[{\citenamefont{Frolov and Pen}(2003)}]{Frolov:2003dk}
\bibinfo{author}{\bibfnamefont{A.~V.} \bibnamefont{Frolov}} \bibnamefont{and}
  \bibinfo{author}{\bibfnamefont{U.-L.} \bibnamefont{Pen}},
  \bibinfo{journal}{Phys. Rev.} \textbf{\bibinfo{volume}{D68}},
  \bibinfo{pages}{124024} (\bibinfo{year}{2003}).
%% \eprint{gr-qc/0307081}.

\bibitem[{\citenamefont{Hamade and Stewart}(1996)}]{Hamade:1995ce}
\bibinfo{author}{\bibfnamefont{R.~S.} \bibnamefont{Hamade}} \bibnamefont{and}
  \bibinfo{author}{\bibfnamefont{J.~M.} \bibnamefont{Stewart}},
  \bibinfo{journal}{Class. Quant. Grav.} \textbf{\bibinfo{volume}{13}},
  \bibinfo{pages}{497} (\bibinfo{year}{1996}).
%%% \eprint{gr-qc/9506044}.
%%-------------------------------------


\bibitem[{\citenamefont{Eto and Sakai}(2003)}]{Eto:2003bn}
\bibinfo{author}{\bibfnamefont{M.}~\bibnamefont{Eto}} \bibnamefont{and}
  \bibinfo{author}{\bibfnamefont{N.}~\bibnamefont{Sakai}},
  \bibinfo{journal}{Phys. Rev.} \textbf{\bibinfo{volume}{D68}},
  \bibinfo{pages}{125001} (\bibinfo{year}{2003}).
%%% \eprint{hep-th/0307276}.



%%-----------------------------------------------
\bibitem{Kudoh:2003xz}
  H.~Kudoh, T.~Tanaka and T.~Nakamura,
  %``Small localized black holes in braneworld: Formulation and numerical
  %method,''
  Phys.\ Rev.\  D {\bf 68}, 024035 (2003);
  H.~Kudoh,
  %``6-dimensional localized black holes: Numerical solutions,''
  Phys.\ Rev.\  D {\bf 69}, 104019 (2004).
%%-----------------------------------------------

  \bibitem{Gibbons:2006ge}
  G.~Gibbons,~K.~Maeda~and Y.~Takamizu,
  %``Fermions on colliding branes,''
  hep-th/0610286.

  
  \bibitem{Griffiths}
  J.~B.~Griffiths,
  {\it Colliding Plane Waves in General Relativity}
  (Oxford University Press, Oxford, 1991).



\bibitem{Hawking:1982ga}
  S.~W.~Hawking, I.~G.~Moss and J.~M.~Stewart,
  %``Bubble Collisions In The Very Early Universe,''
  Phys.\ Rev.\  D {\bf 26}, 2681 (1982).
  %%CITATION = PHRVA,D26,2681;%%




\end{thebibliography}
%%\input{referenceBH_200701.tex} 
%%%%----------------------------------------------------

\end{document}